\documentclass[10pt,a4paper]{article}
\pdfoutput=1
\usepackage[english]{babel}
\usepackage{longtable}
\usepackage{amsmath,amsfonts,amssymb,bm}\interdisplaylinepenalty=2500

\newcommand{\unit}[1]{\ensuremath{\mathrm{#1}}}

\usepackage{graphicx,color,rotating,epstopdf}
\def\PrintGraphicFileName{1}			
\newcommand{\namedgraphics}[3]{
	\parbox{#3}{%
	\ifnum\PrintGraphicFileName>0\rotatebox{90}{\smash{\ttfamily\scriptsize\raisebox{0.8em}{#2}}}\fi%
	\hspace*{\fill}\includegraphics[scale=#1]{#2}\hspace*{\fill}}}
 \newcommand{\namedcombographics}[2]{
	\parbox{\textwidth}{%
	 \ifnum\PrintGraphicFileName>0\rotatebox{90}{\smash{\ttfamily\scriptsize\raisebox{0.8em}{#2}}}\fi%
	\hspace*{\fill}\scalebox{#1}{
	\ifnum\pdfoutput=0\input{\LatexGraphicsPath#2.pstex_t}
	\else\input{\LatexGraphicsPath#2.pdftex_t}\fi}\hspace*{\fill}}}
\newcommand{\namedlatexgraphics}[2]{
	\parbox{\textwidth}{%
	 \ifnum\PrintGraphicFileName>0\rotatebox{90}{\smash{\ttfamily\scriptsize\raisebox{0.8em}{#2}}}\fi%
	\hspace*{\fill}\scalebox{#1}{
	\input{\LatexGraphicsPath#2.latex}}\hspace*{\fill}}}

\title{DFB laser contribution to phase noise in an optoelectronic microwave oscillator}
\author{K. Volyanskiy\thanks{Also with the St.\ Petersburg State University of Aerospace Instrumentation, Russia.}~\thanks{Corresponding author, e-mail vol.kirill@gmail.com},
Y. K. Chembo,
L. Larger,
E. Rubiola\\
\small web page \texttt{http://rubiola.org}
\\[4em]\includegraphics[width=0.35\textwidth]{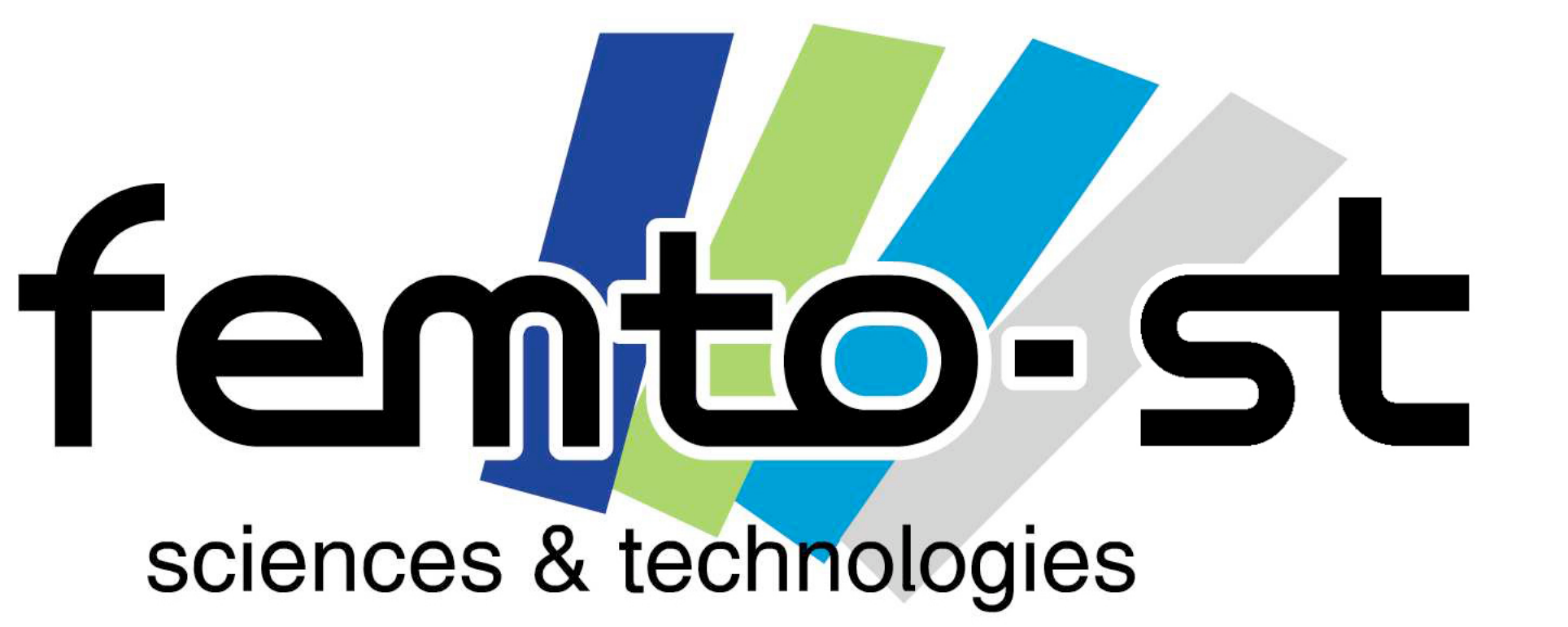}\\[0.5em]
\small FEMTO-ST Institute\\[-0.5ex]
\small CNRS and Universit\'e de Franche Comt\'e, 
\small Besan\c{c}on, France\\[1.5em]}
\date{\small\today}
\def\myheaders{K. Volyanskiy \& al., Laser contrib.\ to OEO noise\hfill\today\quad}
\begin{document}
\maketitle

\begin{abstract}
The optoelectronic fiber delay line microwave oscillator offers attractive and large potential for numerous applications in high speed fiber optics communication, space, and radar systems. Its relevant features are very low phase noise and wide-range tunability, while operating at frequencies of tens of GHz. We consider the contribution of the DFB laser to the oscillator phase noise. Low frequency spectra of wavelength fluctuation and RIN are measured and analyzed.  As a conclusion, the oscillator phase noise can be improved by proper choice of optic and optoelectronic components.
\end{abstract}

\clearpage
\tableofcontents
\clearpage

\section{Introduction}
\begin{figure}[b]
\centering\namedgraphics{0.9}{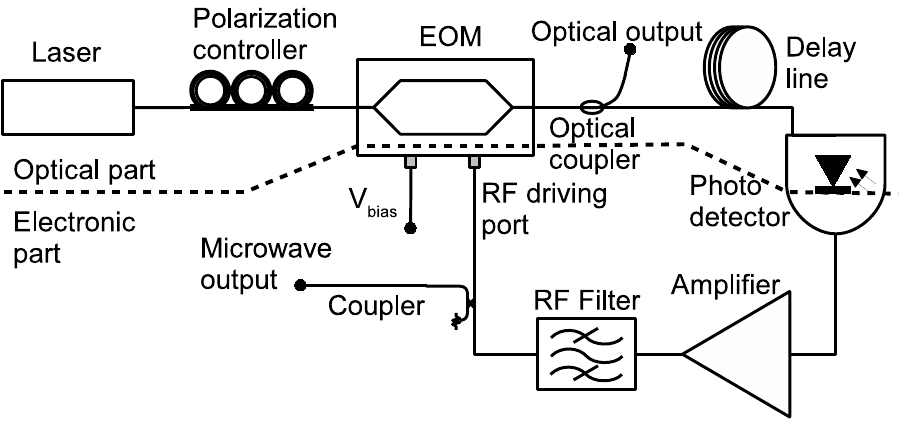}{\columnwidth}
\caption{Basic architecture of the oscillator.}
\label{fig:BA}
\end{figure}

The optoelectronic delay-line microwave oscillator (OEO), shown in Fig.~\ref{fig:BA}, consists of a DFB laser, a high speed electro-optical intensity modulator (EOM), an optical fiber as the delay line, a fast photodetector, a mode selection microwave filter, and an amplifier.

The main reasons to use the fiber carrying a modulated optical beam to implement the microwave delay are (1) the long achievable delay, due to the low loss (0.2 dB/km typical) of the fiber, (2) the wide bandwidth of the delay, (3)  the low background noise, and (4) the low thermal sensitivity of the delay.  The latter has a typical value of $6.85{\times}10^{-6}$/K, a factor of 10 better than the sapphire dielectric cavity.
These features enable the implementation of high spectral purity oscillators and of high-sensitivity instruments for the measurements of phase noise.  In both cases, the optical bandwidth turns into wide-range microwave tunability at virtually no cost in terms of phase noise.

The laser amplitude and phase noise contribute to the OEO phase noise directly or through interaction with other OEO components.  We show this contribution in the case of two telecommunication DFB lasers.

\section{OEO phase noise}\label{ssec:ddl-oscillator-noise}
The OEO under study is organized in a single-loop architecture (Fig.~\ref{fig:BA}). The oscillation loop consists of: a continuous-wave semiconductor laser of optical power $P_{opt}$; a wideband integrated optics LiNbO$_3$ Mach-Zehnder (MZ) modulator characterized by a
half-wave voltage $V_{\pi} = 4$ V; a thermalized 4 km fiber performing a time delay of $T = 20$ \unit{\mu s} on the microwave signal carried by the optical beam (the corresponding free spectral range is $1/T = 50$ kHz); a fast photodiode with a conversion factor $\rho$; a narrow band microwave radio-frequency (RF) filter, of central frequency $F_0 = \Omega_0/2\pi = 10$ GHz, and $-3$ dB bandwidth of $\Delta F = \Delta\Omega/2\pi = 50$ MHz; a microwave amplifier with gain $G$. All optical and electrical losses are gathered in a single attenuation factor $\kappa$. The dynamics of OEO microwave oscillation can be described in terms of the dimensionless variable $x(t) = \pi V(t)/2V_{\pi}$ that obeys \cite{Y.K.2007}
\begin{align}
x+\tau \frac{dx}{dt}+\frac{1}{\theta}\int^{t}_{t_0} x(s)ds = \beta cos^2 [x(t-T)+\phi],
\end{align}
where $\beta = \pi \kappa GP_{opt}\rho R_{ph}/2V_{\pi}$ is a normalized loop gain, $R_{ph}$ is a photodiode current-voltage conversion resistance, $\phi = \pi V_B/2V_{\pi}$ is the Mach-Zehnder offset phase, while $\tau = 1/\Delta \Omega$ and $\theta = \Delta \Omega / \Omega_0^2$ are the characteristic timescale parameters of the bandpass filter.

The dynamics of slowly varying envelope ${\cal A} = A(t)e^{i\psi(t)}$ of the microwave $x(t)$ can be expressed by the following linearized equation \cite{Y.K.2008}
\begin{align}
\dot{\cal A} = -\mu e^{i\vartheta}{\cal A}+ \mu e^{i\vartheta}[1 + \eta_m (t)]{\cal A}_T + \mu e^{i\vartheta} \zeta_a(t),
\end{align}
where $\eta_m$ is a dimensionless multiplicative noise ($\eta_m(t) = \delta \gamma (t)/\gamma$, $\gamma = \beta \sin 2\phi$ is a normalized loop gain with taking into account the Mach-Zehnder offset phase), $\zeta_a$ is an additive noise (a complex Gaussian white noise), $\vartheta\cong1/(2Q)$, $Q$ is the quality factor of the bandpass filter, $\mu \cong \Delta \Omega /2$.

Using the It\^o rules of stochastic calculus \cite{Y.K.2008}, we derive the following time-domain equation for the phase dynamics
\begin{align}
\dot{\psi}= -\mu(\psi - \psi_T) + \frac{\mu}{2Q}\eta_m(t) + \frac{\mu}{\left|{\cal A}_0\right|}\xi_{a,\psi}(t),
\end{align}
where $\xi_{a,\psi}(t)$ is a real Gaussian white noise of correlation $\left\langle{\xi}_a(t) {\xi}_a(t') \right\rangle = 2D_a \delta(t-t')$ (same variance as $\xi_a(t)$). We can add $\varsigma_{\psi}(t)$ to take into account the contribution of laser frequency noise through the delay line dispersion:
\begin{align}
\dot{\psi}= \mu\left(\psi_T - \psi + \varsigma_{\psi}(t) + \frac{\eta_m(t)}{2Q} + \frac{\xi_{a,\psi}(t)}{\left|{\cal A}_0\right|}\right).
\label{eq:dpsi}
\end{align}

We can use Eq. (\ref{eq:dpsi}) to obtain the Fourier spectrum $\Psi(\omega)$ of the phase $\psi(t)$, and then its power density spectrum following \cite{Y.K.2008}
\begin{align}
\left|\Psi(\omega)\right|^2=\left|\mu\frac{\frac{\tilde{\eta}_m(\omega)}{2Q}+\frac{\sqrt{2D_a}}{|{\cal A}_0|}+\varsigma_{\psi}(\omega)}{i\omega+\mu\left[1-e^{-i\omega T}\right]}\right|^2 .
\label{eq:Psi}
\end{align}

The diffusion constant $D_a$ can be calculated using the characteristics of OEO components. The noise power at the amplifier output is caused by laser RIN at 10 GHz, the thermal noise, and the shot noise; it can be expressed as:
\begin{align}
P_0 =\left[N_{RIN}I_{ph}^2R_{eq} + FkT_0 + 2eI_{ph}R_{ph}\right] \frac{G\Delta F}{2} ,
\end{align}
where $G$ is the amplifier gain, $F$ is the noise figure of amplifier, $T_0 = 295$K is the room temperature, $k$ is the Boltzmann constant, $e$ is the electron charge, $I_{ph}$ is the photodiode current, $\Delta F$ is the RF filter bandwidth. Here we take into account only high frequency laser RIN. 

$D_a$ can be determined \cite{Y.K.2008} as
\begin{align}
D_a = \frac{P_0\pi R}{4V_{\pi}^2\Delta F} ,
\end{align}
where $R$ is the output impedance (in our case, $R = 50$ \unit{\Omega}).

\section{RIN and frequency noise of DFB lasers}
\begin{figure}[t]
\centering\namedgraphics{0.8}{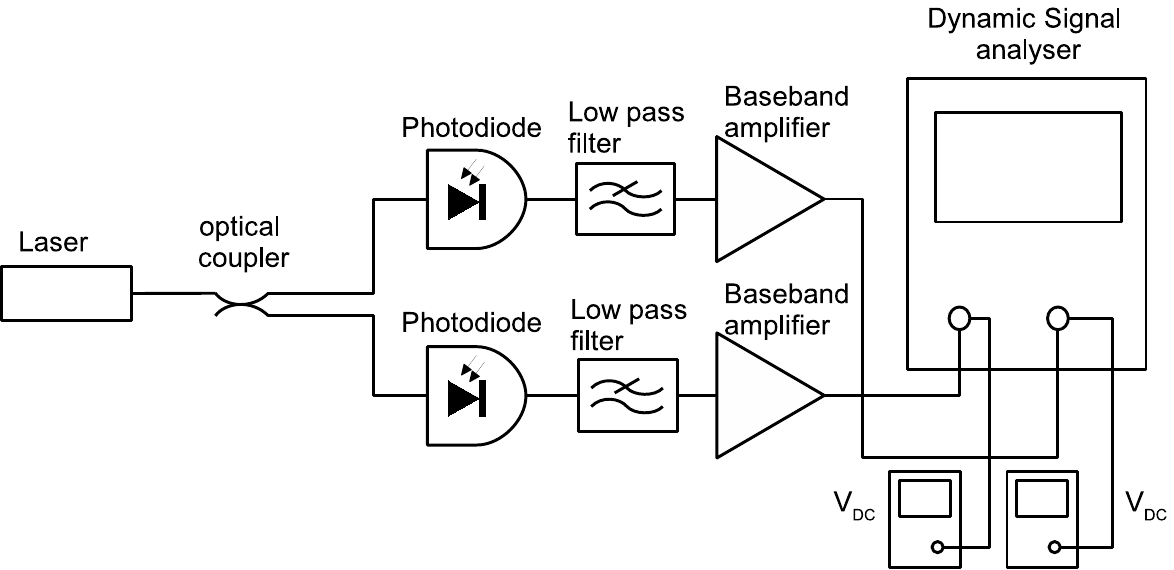}{\columnwidth}
\caption{RIN measurement bench.}
\label{fig:RINMeas}
\end{figure}

To evaluate the contribution of DFB lasers to the phase noise of OEO we should also measure their low frequency relative intensity noise (RIN) and frequency noise. The following measurement scheme (Fig.~\ref{fig:RINMeas}) was used to measure the low frequency RIN.

\begin{figure}[t]
\centering\namedgraphics{0.50}{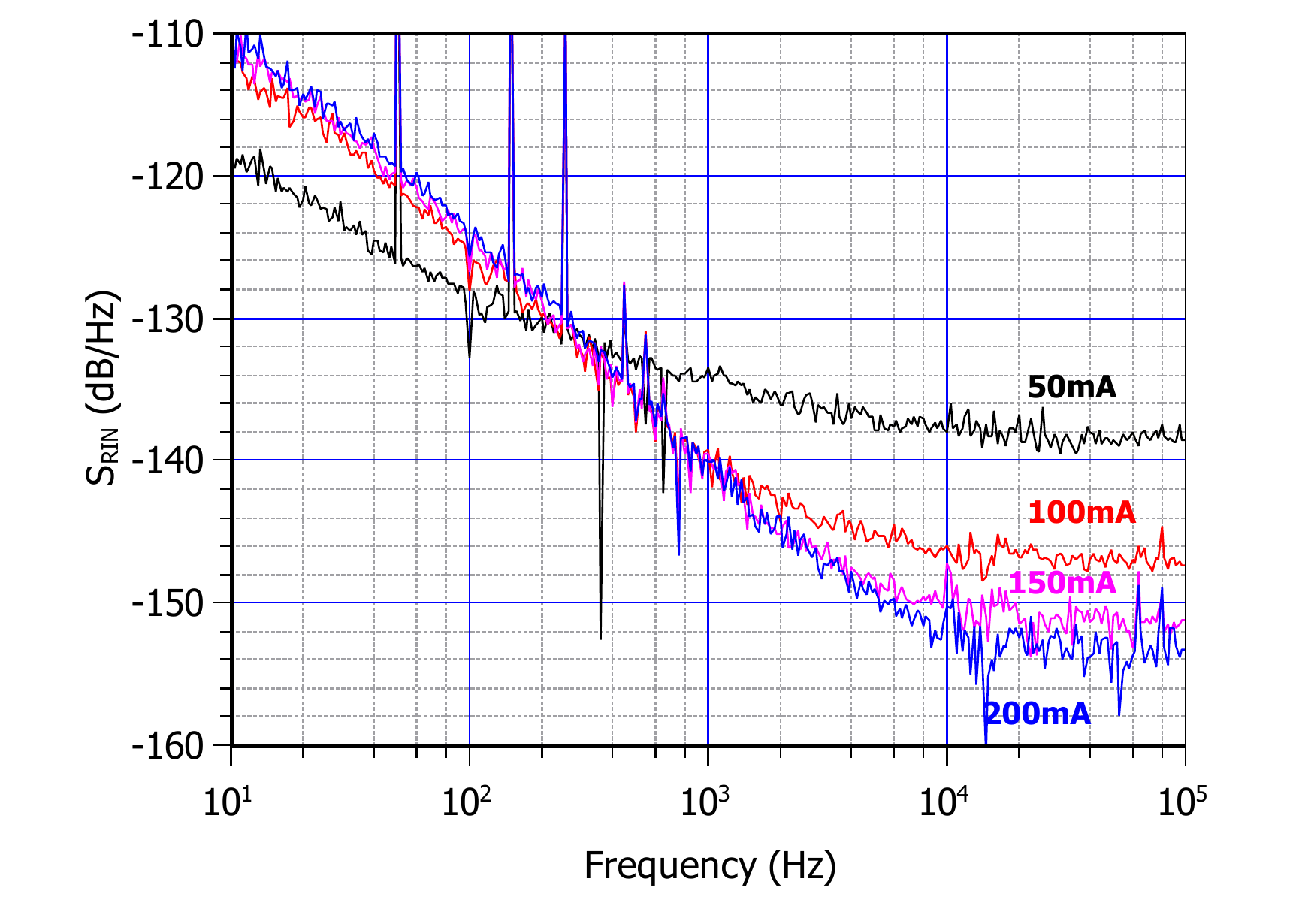}{\columnwidth}
\caption{Laser EM4 RIN.}
\label{fig:EM4RIN}
\end{figure}

\begin{figure}[t]
\centering\namedgraphics{0.50}{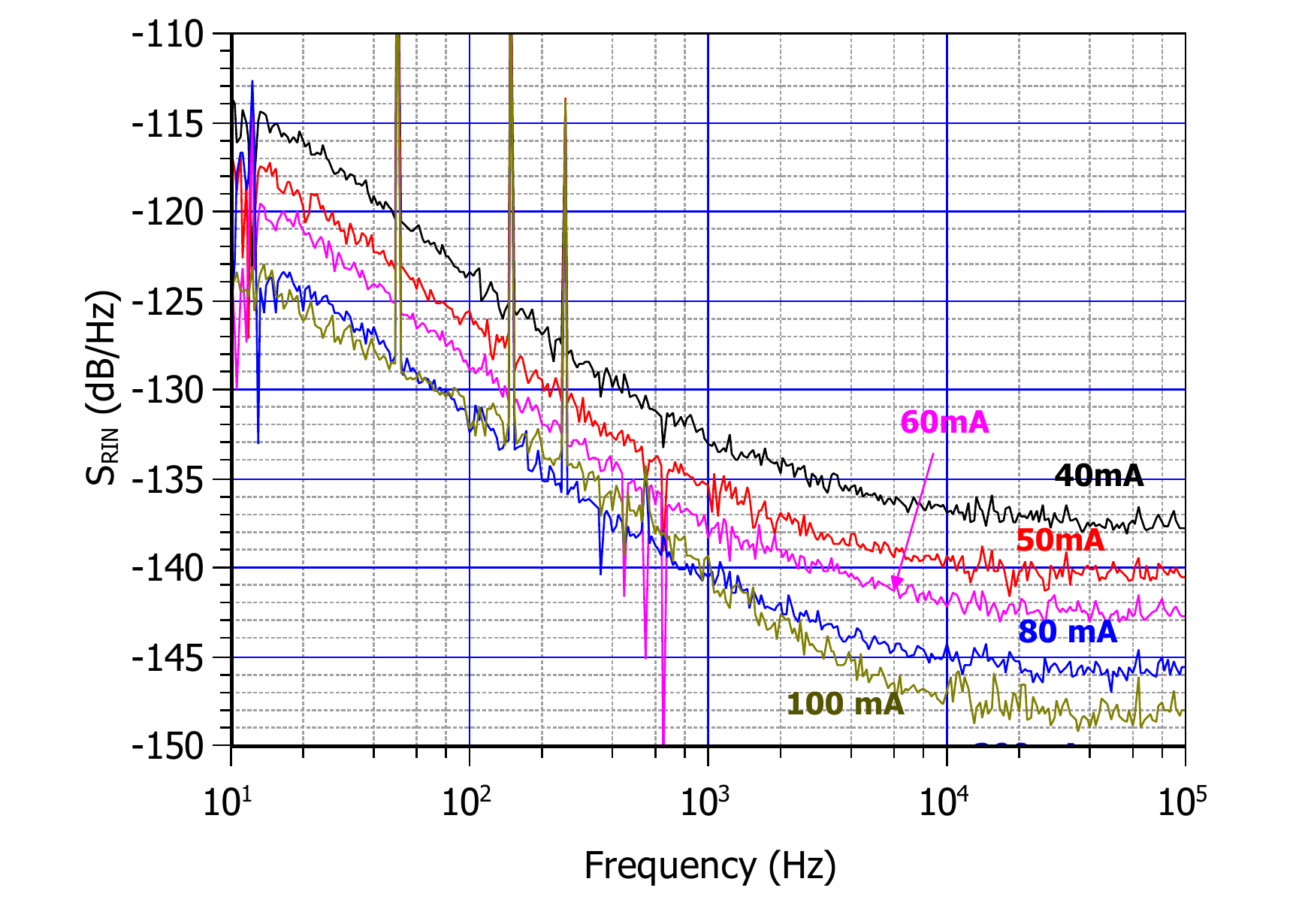}{\columnwidth}
\caption{Laser CQF935 RIN.}
\label{fig:CQFRIN}
\end{figure}

In such a way we could measure the low frequency RIN in a spectral range up to 100 kHz. We used a cross-correlation method to eliminate the phase noise of the photodiodes and the amplifiers. The results are presented in Figs. \ref{fig:EM4RIN} and \ref{fig:CQFRIN}.

\begin{figure}[t]
\centering\namedgraphics{0.8}{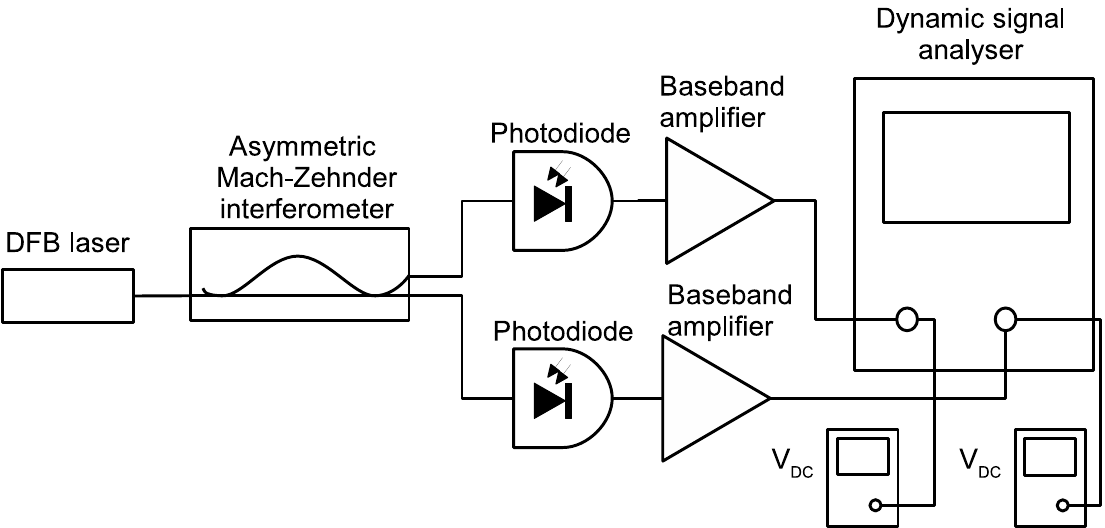}{\columnwidth} 
\caption{Frequency noise measurement bench.}
\label{fig:FFMeas}
\end{figure}

The EM4 laser is more powerful (up to 50 mW, 450 mA) than the CQF935 laser (up to 20 mW, 100 mA), but it has a slightly higher RIN. The dependence of RIN on laser current is similar for both lasers.
A similar scheme (Fig.~\ref{fig:FFMeas}), but with an asymmetric Mach-Zehnder interferometer serving as an optical frequency detector was used to measure the laser frequency noise.

Power spectral density (PSD) was chosen as output of the dynamical signal analyser, and a correct calibration of noise could be achieved with an adequate conversion coefficient. The latter can be obtained from the analysis of the optical power transfer function of the interferometer
\begin{equation}
K(f) = \frac{1}{2}+\frac{\cos(2\pi f T_D)}{2},
\end{equation}
where $T_D$ is the differential delay of the asymmetric MZ interferometer, $f$ - optical frequency.
\begin{equation}
\frac{\partial K(f)}{\partial f} = -\pi T_D\sin(2\pi f T_D).
\end{equation}
For a given laser wavelength, the interferometer unbalancing $T_D$ can be finely tuned (through thermal effect in the device or wavelength of laser), so that the laser frequency noise detector operates as a linear detector $(\sin(2\pi f T_D) \simeq \pm 1)$. In this case the transfer function can be represented as 
\begin{equation}
K(f) \approx K_0 + \delta f \cdot \left(\frac{\partial K(f)}{\partial f}\right)_{f_0}.
\end{equation}

\begin{figure}[t]
\centering\namedgraphics{0.50}{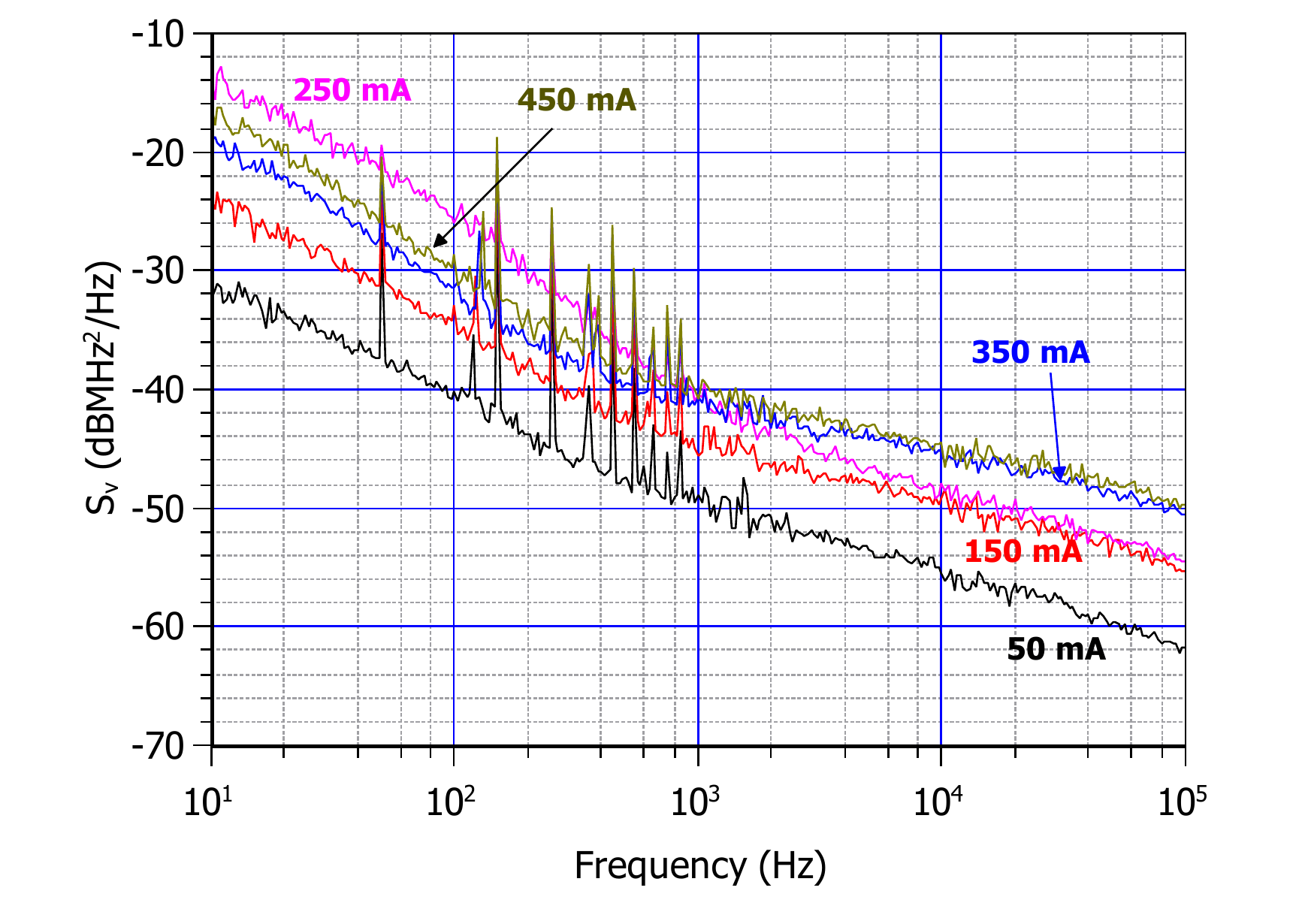}{\columnwidth}
\caption{EM4 optical frequency noise.}
\label{fig:EM4FF}
\end{figure}

\begin{figure}[t]
\centering\namedgraphics{0.50}{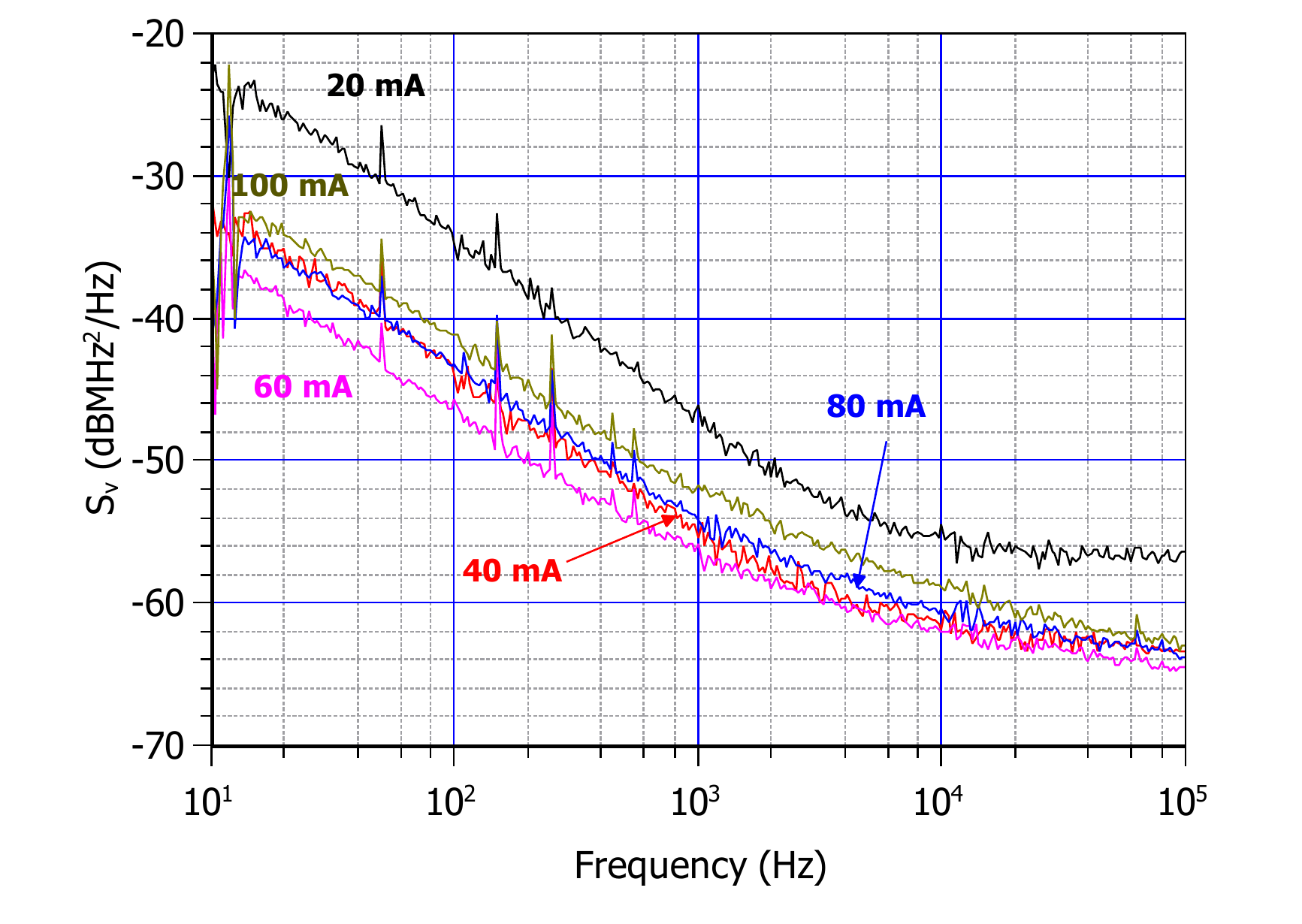}{\columnwidth}
\caption{CQF935 optical frequency noise.}
\label{fig:CQFFF}
\end{figure}

So the conversion factor for frequency noise is 
\begin{equation}
C_f = \frac{\delta f}{\delta V} = \frac{1}{\frac{\partial K(f)}{\partial f}P_0 \rho R_{ph} G} = \frac{1}{\pi P_0 T_D \rho R_{ph} G},
\end{equation}
where $P_0$ is the lasing power, $\rho$ is the photodiode conversion factor, $R_{ph}$ is the photodiode current-voltage conversion resistance, $G$ is the amplifier gain. The constant voltage at the output of amplifier can be defined as
\begin{equation}
	V_{DC} = \frac{P_0 \rho R_{ph} G}{2}.
\end{equation}
Therefore the conversion factor is simplified as follows
\begin{align}
C_f = \frac{1}{2\pi T_D V_{DC}},
\end{align}
where $T_D$ is the differential delay ($T_D = 402.68$ ps in our case).
The results are presented in Figs. \ref{fig:EM4FF} and \ref{fig:CQFFF}.

We can see from the diagrams that the EM4 laser has significantly higher frequency noise. The dependence of frequency noise on laser current for EM4 differs from that of CQF935.

\section{Contribution to the phase noise of OEO}
We used low phase noise amplifiers ($-154$ dBc/Hz at 1 kHz, gain 22 dB) in our architecture. One such amplifier in the loop provided necessary gain for OEO functioning at high power mode of the EM4 laser and two such amplifiers provided necessary gain at low power mode of the EM4 laser or at using the CQF935 laser. The typical flicker coefficient of a InGaAs p-i-n photodetector is of $10^{-12}$ \unit{rad^2/Hz} ($-120$ \unit{dBrad^2/Hz}) \cite{Shieh1998,E.2006,Shieh2005}. Other components have supposedly lower magnitudes of the phase noise. We will consider the laser contribution to the phase noise of an OEO of the architecture in Fig. \ref{fig:BA}, with two different DFB lasers at different operating powers. Both lasers have 1550 nm wavelength.

The low frequency RIN (Figs. \ref{fig:EM4RIN} and \ref{fig:CQFRIN}) can be represented by multiplicative noise $\eta_m(\omega)$ in Eq. (\ref{eq:Psi}). In the most cases, the quality factor of the RF filter is high ($Q = 200$ in our case) and therefore the low frequency RIN can be neglected. Laser RIN for microwave frequencies (10 GHz in our case) is about $-160$ dB/Hz according to data sheets of the lasers.

\begin{figure}[t]
\centering\namedgraphics{0.50}{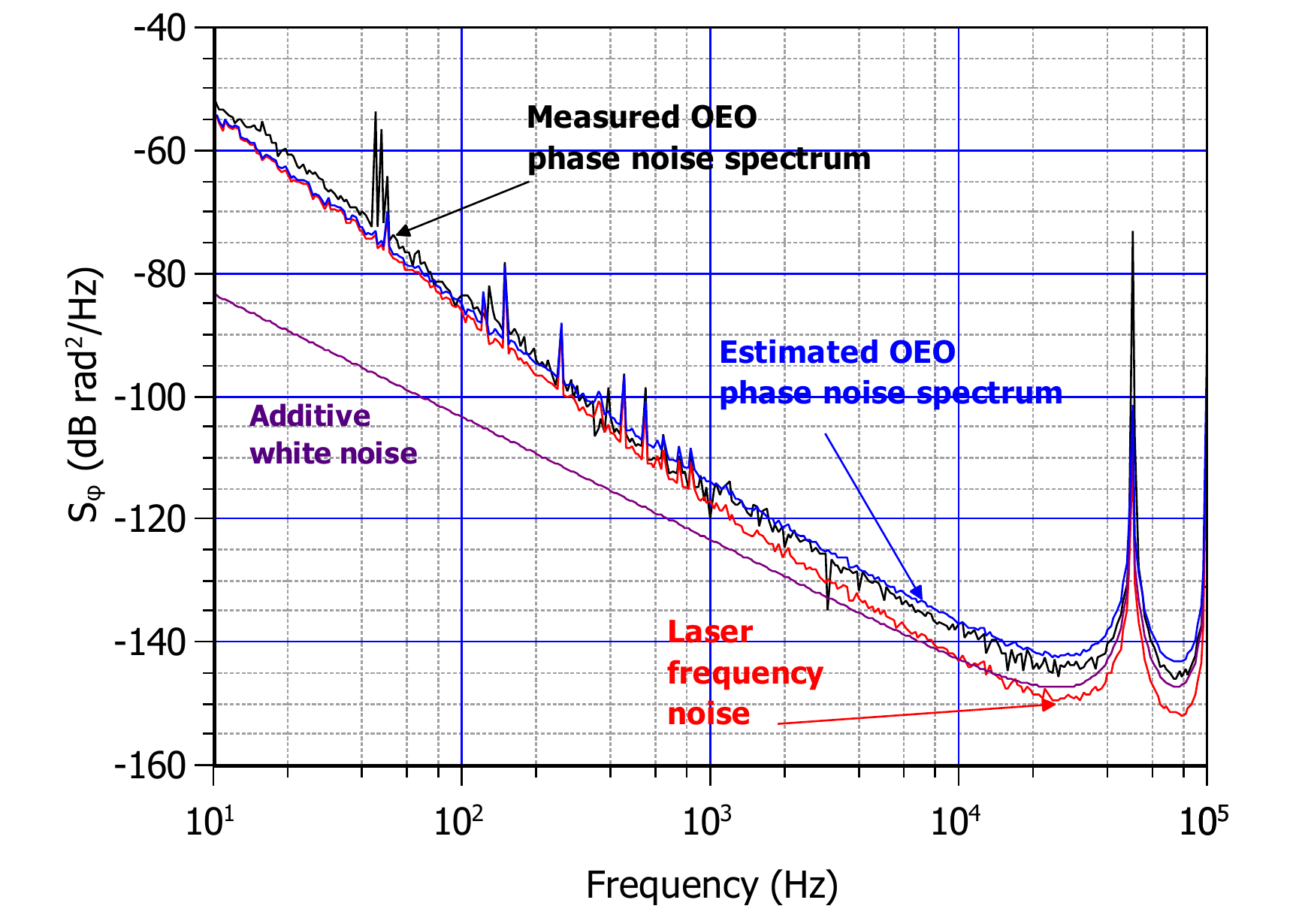}{\columnwidth}
\caption{The phase noise of OEO with the EM4 laser at laser current 200 mA. Two low phase noise microwave amplifiers (G = 22 dB) are used.}
\label{fig:OEOEM4}
\end{figure}

\begin{figure}[t]
\centering\namedgraphics{0.50}{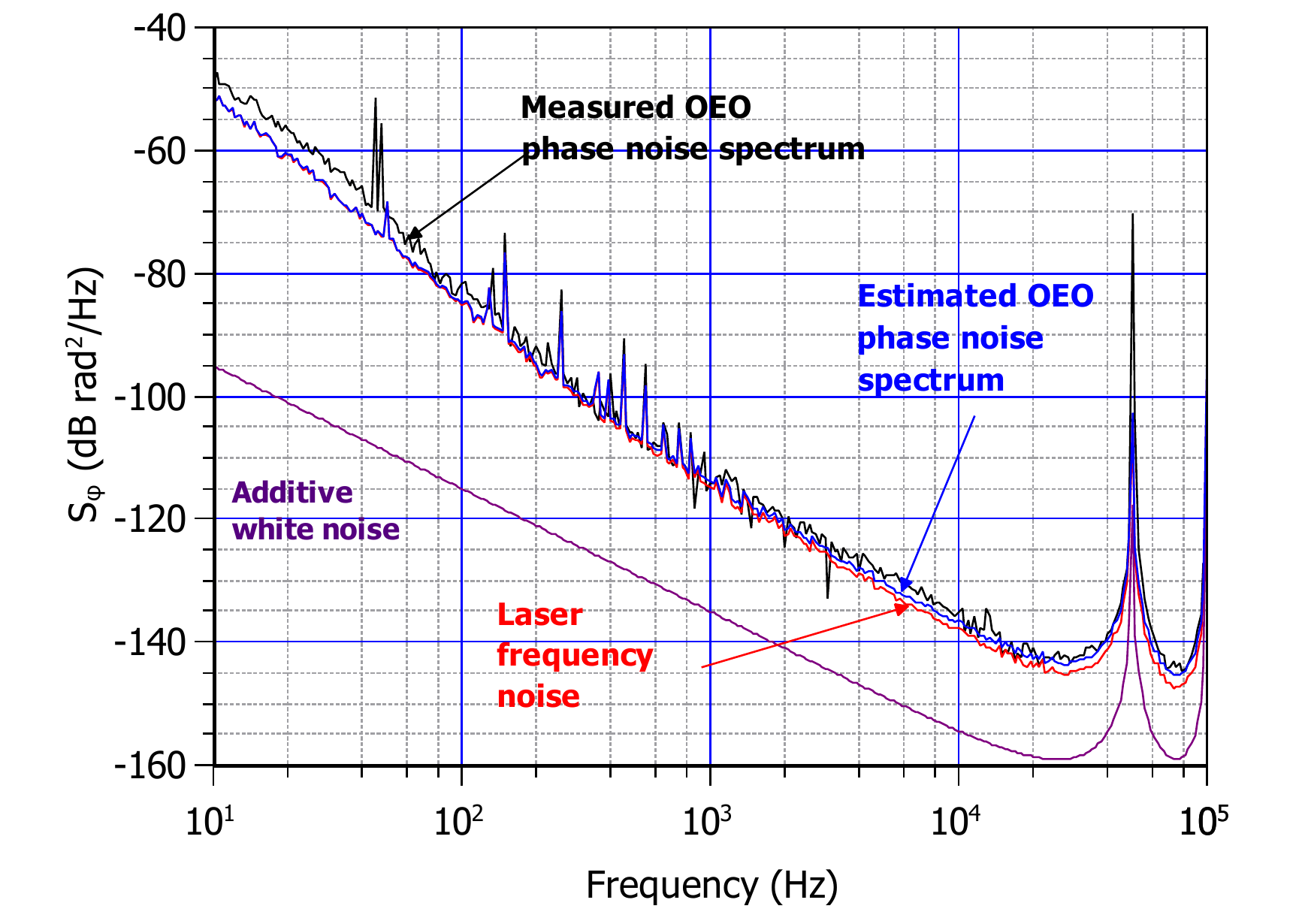}{\columnwidth}
\caption{The phase noise of OEO with the EM4 laser at laser current 400 mA. One low phase noise microwave amplifier (G = 22 dB) is used.}
\label{fig:OEOEM4l2}
\end{figure}

\begin{figure}[t]
\centering\namedgraphics{0.50}{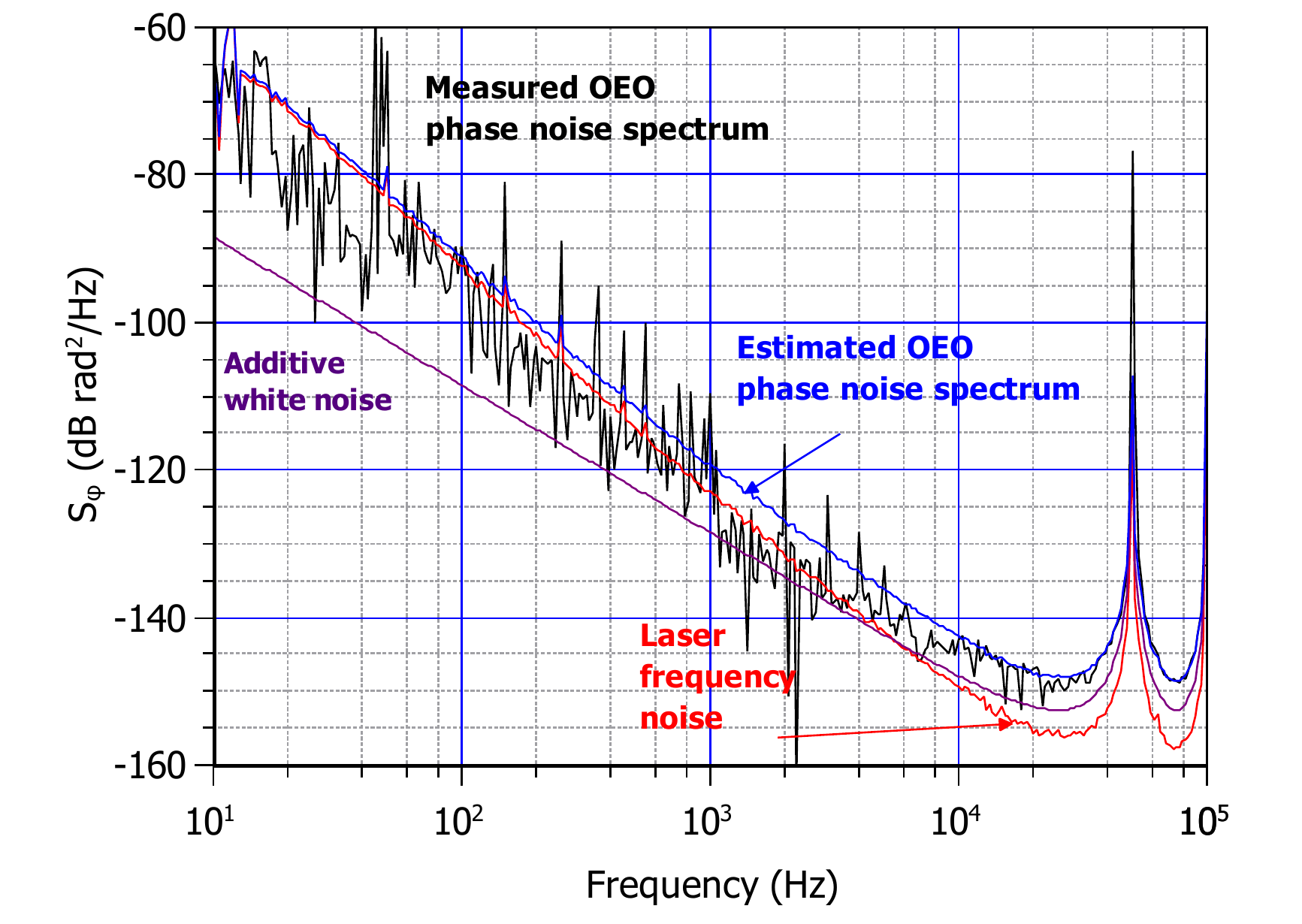}{\columnwidth}
\caption{The phase noise of OEO with the CQF935 laser at laser current 100 mA. Two low phase noise microwave amplifiers (G = 22 dB) are used.}
\label{fig:OEOCQF}
\end{figure}

The optical delay line has an optical length that is defined by its physical length and its refractive index, which depends on optical frequency. This dependence can be calculated with a dispersion constant, which can be found in a data sheet of particular optical fiber for given wavelength. Such dispersion constant for optical fiber SMF-28e that we used in our experiments is $D_\lambda = 18$ \unit{ps/(km\cdot nm)} at 1550 nm. 
Thus, to calculate the phase noise caused by the frequency noise over 4 km of optical fiber SMF-28e, we used the following conversion factor
\begin{align}
C_{\psi} = \frac{2\pi D_{\lambda}L \lambda_0^2}{c \: T_0 \: n},
\end{align}
where $D_{\lambda}$ is the optical fiber dispersion (17 ps/nm/km for SMF-28 optical fiber), $L$ is the delay line length (4 km), $c$ is the light speed in vacuum, $T_0$ is the microwave oscillation period, $n$ is the optical fiber refraction index (1.46).
Equation (\ref{eq:Psi}) was used to estimate the OEO phase noise, taking into account the laser RIN at 10 GHz and its optical frequency noise, the thermal white noise, and the photodiode shot noise.
The comparison of frequency noise contribution and additive white noise contribution is shown in Figs. \ref{fig:OEOEM4} - \ref{fig:OEOCQF}.

As we can see the major part of the phase noise in this configuration is created by the laser frequency noise through the delay line dispersion. It is possible to use dispersion shifted optical fiber with zero dispersion at the lasing wavelength in order to strongly limit this phase noise contribution. It should thus considerably decrease the OEO phase noise level. At the same time, it is preferable to use the architecture with one amplifier as we can possibly achieve almost $-160$ \unit{dB rad^2/Hz} at the frequency of about 25 kHz and $-95$ \unit{dB rad^2/Hz} at the frequency of about 10 Hz (Fig. \ref{fig:OEOEM4l2}).

Here we have a good agreement between prediction and experimental results. Remaining little discrepancies can be caused by errors in measuring the values that are used for prediction, like the oscillation power, the losses in microwave circuits, the laser RIN.

\section{Conclusion}
This article presents a theoretical and experimental study of the phase noise in OEOs. We have used the phase noise expression derived with using the Langevin formalism \cite{Y.K.2008} and added laser frequency noise contribution. We have found a good agreement between the main predictions of the model and the experimental results. Some suggestions on improving the OEO architecture were proposed. The article enables to achieve better understanding of phase noise components in OEO.

\def\bibfile#1{#1}
\bibliographystyle{abbrv}
\bibliography{\bibfile{bibliog}}
\addcontentsline{toc}{section}{References}

\end{document}